\documentclass[a4paper,12pt]{article}
\usepackage{amsfonts}
\usepackage{amsmath,amssymb}
\usepackage[utf8]{inputenc}
\usepackage[T1]{fontenc}
\usepackage{lmodern}
\usepackage[left=2cm,right=1.5cm,top=1.5cm,bottom=1.5cm,bindingoffset=0cm]{geometry}

\numberwithin{equation}{section}
\usepackage{amsthm}
\usepackage[hidelinks]{hyperref}
\newtheorem{theorem}{Theorem}[section]
\newtheorem{proposition}[theorem]{Proposition}
\newtheorem{remark}[theorem]{Remark}
\newcommand{\D}{\partial}
\allowdisplaybreaks[2]
\date{}
\begin{document}
	
	\title{Remarks on the complete integrability of the $A_\pm$ equation}
	\author{\textbf{Yarema A. Prykarpatskyy} \\
	Department of Applied Mathematics, University of Agriculture in Kraków, \\
	Aleja Mickiewicza 21, Kraków, 31-120, Poland  \\
	Institute of Mathematics of NAS of Ukraine, 3 Tereschenkivska st.,\\ Kyiv-4, 01024, Ukraine
	}
\date{}
	\maketitle
	
	\begin{abstract}
		We study the integrability properties of Alexeyev's $A_\pm$ equation by means of the gradient-holonomic approach. The Lax--Noether equation is investigated, and the first coefficients of its asymptotic solution are explicitly calculated. For the $A_+$ equation, two compatible formal Hamiltonian operators are constructed directly by the gradient-holonomic scheme using the conservation laws, and the corresponding recursion operator is obtained. A logarithmic point transformation identifies the equation on the nonvanishing branch with the Calogero--Degasperis--Fokas family. A differential substitution into the Korteweg--de Vries equation provides an independent construction of infinitely many local conservation laws, with functional independence of arbitrary finite initial subfamilies on suitable open sets. A combined Lax representation is established on the positive periodic branch. The distinction between the formal bi-Hamiltonian representation and its periodic realization is discussed.
	\end{abstract}
	
	
	\section{Introduction}
	
	In the framework of the multidimensional superposition principle developed in \cite{Alex2003}, the following nonlinear equation was constructed in \cite{Alex00}:
	\begin{equation}
		\label{eq00}
		(v^3)_t-2(v^3)_{xxx}+3v_x(18v_{xx}v+v_x^2+3p+3sv^4)=0,
	\end{equation}
	and named the $A_\pm$ equation. Here $v=v(x,t)$, $s\in\{1,-1\}$ denotes the choice of sign and $p\in\mathbb R$ is a constant parameter. In the notation of \cite[(3.37)]{Alex00}, the constant term in the parentheses is $\pm3p_{\rm A}^2$; we absorb this sign into our parameter $p=sp_{\rm A}^2$. Before fixing the spatial period, the nonzero parameter can be normalized to $\pm1$ by suitable scaling. The two sign choices are connected, over the complex numbers, by $(v,p,s)\mapsto(iv,-p,-s)$; the equation is also invariant under $v\mapsto-v$.
	
	In the paper \cite{Alex01}, the integrability of the equation (\ref{eq00}) was investigated by means of the Painlev\'e property and the Hirota three-soliton test. Multiple bell-shaped soliton and breather solutions were also obtained. These results indicate that the equation is integrable by the inverse scattering transform. The author reported that his initial search for a second-order AKNS-type inverse scattering problem was unsuccessful, suggesting the need to consider a broader class of spectral problems. The earlier paper \cite{Alex2003} developed the soliton invariant-manifold method and illustrated it, in particular, for the KdV and mKdV equations; the $A_\pm$ equation itself was introduced subsequently in \cite{Alex00}.
	
	In the present paper we use the gradient-holonomic scheme to construct conservation laws and a formal bi-Hamiltonian structure for the $A_+$ equation. Independently, we establish the existence of an infinite hierarchy of local conservation laws for the $A_\pm$ equation, with functional independence in the finite-family sense specified below. We also show that, on the nonvanishing branch, equation (\ref{eq00}) is point-equivalent to a member of the Calogero--Degasperis--Fokas (CDF) family. The CDF equation and its Miura transformations and Lax representations have an earlier literature \cite{CalDeg,Fokas,Hlav,Cao,Hone}. The purpose of establishing this connection here is to obtain a Lax representation and an independent construction of a hierarchy of conservation laws, rather than to claim novelty for the classical CDF hierarchy itself.
	
	The paper is structured as follows. In Section~\ref{ConsLaws} we investigate conservation laws by the gradient-holonomic scheme and construct two compatible formal Hamiltonian operators and their recursion operator. In Section~\ref{sec:CDF} we establish the connection with the CDF and KdV equations, derive a Lax representation and construct an infinite family of local conserved functionals whose finite initial subfamilies are functionally independent on suitable open sets. The final section summarizes the results and the remaining questions.
	
	\section{Conservation laws}\label{ConsLaws}

	To proceed, we use the following obvious substitution $v=u^{\frac13}$. The equation (\ref{eq00}) will be rewritten as follows:
	\begin{equation}
		\label{eq01}
		u_t=2u_{xxx}+\frac{u_x}{9}\left(-54u_{xx}u^{-1}+35u_x^2u^{-2}-27pu^{-2/3}-27su^{2/3}\right).
	\end{equation}
	For the simplicity of the analysis we first take only the ``$+$'' sign, i.e., $s=1$.
	
	For studying the integrability properties of dynamical system (\ref{eq01}) we will apply the gradient-holonomic scheme devised in \cite{AK00,PM,BPS}. To proceed, let us assume that the dynamical system (\ref{eq01}) is given in the form $u_t=K[u]$ on the infinite-dimensional periodic manifold
	\[
	M=\{u\in C^\infty(\mathbb R/2\pi\mathbb Z,\mathbb R):u>0\},
	\]
	as a smooth Fr\'echet differentiable vector field $K:M\to T(M)$. The positive branch is chosen to avoid singular coefficients at $u=0$; the negative branch is obtained by $v\mapsto-v$.
	Following the approach, we first describe the asymptotic solutions to the Lax--Noether equation \cite{Lax,BPS}
	\begin{equation}\label{eq02}
		\varphi_t+K'^*[u]\varphi=0,
	\end{equation}
	where the prime is the Fr\'echet derivative and the star denotes the formal adjoint with respect to the bilinear form $(\cdot\mid\cdot)$ on $T^*(M)\times T(M)$, defined as
	\begin{equation}\label{eq03}
		(a\mid b)=\int_0^{2\pi}a(x)b(x)\,dx
	\end{equation}
	for any $a\in T^*(M)$, $b\in T(M)$. Complexification is understood when a complex spectral parameter is used. The linear differential adjoint operator $K'^*[u]:T^*(M)\to T^*(M)$ has the form
	\begin{equation}\label{eq04}
		\begin{aligned}
			K'^*[u]={}&-\frac{16u_xu_{xx}}{3u^2}+\frac{32u_x^3}{9u^3}
			+\left(-\frac{6u_{xx}}u+\frac{u_x^2}{3u^2}+3u^{2/3}+3pu^{-2/3}\right)\partial\\
			&-\frac{6u_x}u\partial^2-2\partial^3,
		\end{aligned}
	\end{equation}
	where $\partial=D_x$. The Lax--Noether equation (\ref{eq02}) allows a formal asymptotic solution
	\begin{equation}\label{eq05}
		\varphi(x,t;\lambda)=\exp[\omega(\lambda)t+\partial^{-1}\sigma(x,t;\lambda)].
	\end{equation}
	Here $\lambda\in\mathbb C$ is a complex parameter. In (\ref{eq05}), $\partial^{-1}$ denotes a local primitive on the universal cover of the circle; its $x$-independent part can depend on $t$ and is chosen consistently with (\ref{eq02}). The formal solution $\varphi$ itself need not be periodic, whereas its logarithmic derivative $\sigma=\varphi_x/\varphi$ has periodic differential-function coefficients. Thus $\int_0^{2\pi}\sigma(x,t;\lambda)\,dx$ is a generating function of the corresponding conservation laws. A genuine inverse of $\partial$ on all periodic functions does not exist; this issue is also important in the Hamiltonian construction below.
	
	The principal part $-2\partial^3$ of (\ref{eq04}) gives the leading dispersion term $2\lambda^3$. We can write the formal solution in the form
	\begin{equation}\label{eq06}
		\varphi(x,t;\lambda)=\exp[2\lambda^3t+\partial^{-1}\sigma(x,t;\lambda)],
	\end{equation}
	where the normalization of the primitive includes any remaining $x$-independent time terms, and
	\begin{equation}\label{eq07}
		\sigma(x,t;\lambda)\simeq\lambda+\sum_{j=0}^{\infty}\sigma_j[u]\lambda^{-j}.
	\end{equation}
	At a constant field $u=1$, for example, the dispersion of an exponential with wave number $k$ is $2k^3-3(1+p)k$; the wave number in (\ref{eq07}) is a formal series in $\lambda$, rather than necessarily $\lambda$ itself.
	
	After substitution of (\ref{eq06}) and (\ref{eq07}) into (\ref{eq02}), using
	\[
	\varphi_x=\sigma\varphi,\quad
	\varphi_{xx}=(\sigma_x+\sigma^2)\varphi,\quad
	\varphi_{xxx}=(\sigma_{xx}+3\sigma\sigma_x+\sigma^3)\varphi,
	\]
	we obtain the generating equation
	\begin{subequations}\label{eq08}
		\begin{equation}\label{eq:recursion-master}
			\begin{aligned}
				2\lambda^3+\partial^{-1}\sigma_t={}&2(\sigma_{xx}+3\sigma\sigma_x+\sigma^3)
				+\frac{6u_x}{u}(\sigma_x+\sigma^2)\\
				&+\left(\frac{6u_{xx}}u-\frac{u_x^2}{3u^2}-3u^{2/3}-3pu^{-2/3}\right)\sigma\\
				&+\frac{16u_xu_{xx}}{3u^2}-\frac{32u_x^3}{9u^3}.
			\end{aligned}
		\end{equation}
		Equating the coefficients of $\lambda^{-j}$ in (\ref{eq:recursion-master}), we obtain the following system of recursion equations:
		\begin{equation}\label{eq:recursion-system}
			\begin{aligned}
				0={}&2\delta_{j,-3}+\partial^{-1}\sigma_{j,t}-2\sigma_{j,xx}
				-6\sum_{m+n=j}\sigma_m\sigma_{n,x}
				-2\sum_{m+n+\ell=j}\sigma_m\sigma_n\sigma_\ell\\
				&-\frac{6u_x}{u}\sigma_{j,x}
				-\frac{6u_x}{u}\sum_{m+n=j}\sigma_m\sigma_n\\
				&+\left(-\frac{6u_{xx}}{u}+\frac{u_x^2}{3u^2}+3u^{2/3}+3pu^{-2/3}\right)\sigma_j\\
				&+\left(-\frac{16u_xu_{xx}}{3u^2}+\frac{32u_x^3}{9u^3}\right)\delta_{j,0},
				\qquad j\geq -3.
			\end{aligned}
		\end{equation}
	\end{subequations}
	Here $m,n,\ell\geq-1$ in all sums, $\delta_{j,k}$ is the Kronecker delta, $\sigma_{-1}=1$ and $\sigma_j=0$ for $j<-1$. The primitive in (\ref{eq:recursion-system}) is taken coefficientwise, with its integration constants fixed by the chosen spectral normalization; the coefficients of $\partial^{-1}\sigma_t$ at positive powers of $\lambda$ are zero for the normalization used here.
	
	For $j=-3$, the system gives $2-2\sigma_{-1}^3=0$. For $j\geq-2$, the three permutations of $(-1,-1,j+2)$ in the cubic sum contribute $-6\sigma_{j+2}$. All other terms contain only coefficients of index at most $j+1$. Thus (\ref{eq:recursion-system}) successively determines $\sigma_{j+2}$ once the primitive is normalized. For example, the equations corresponding to $j=-2,-1$ are
	\[
	\begin{aligned}
		0={}&-6\sigma_0-6\frac{u_x}{u},\\
		0={}&-6\sigma_1-6\sigma_{0,x}-6\sigma_0^2-12\frac{u_x}{u}\sigma_0\\
		&-\frac{6u_{xx}}{u}+\frac{u_x^2}{3u^2}+3u^{2/3}+3pu^{-2/3}.
	\end{aligned}
	\]
	These give $\sigma_0=-u_x/u$ and $\sigma_1=u_x^2/(18u^2)+u^{2/3}/2+pu^{-2/3}/2$.
	Equivalently, differentiating (\ref{eq:recursion-master}) with respect to $x$ eliminates the primitive and gives a conservation-form equation for $\sigma_t$. The recursion gives the successive coefficients formally; their locality at arbitrary order is not assumed here.
	Solving these equations sequentially, we obtain the following generating densities:
	
	\begin{equation}\label{eq09}
		\begin{aligned}
			\sigma_{-1}={}&1,\qquad \sigma_0=-\frac{u_x}{u},\\
			\sigma_1={}&\frac{u_x^2}{18u^2}+\frac{u^{2/3}}2+\frac{p}{2u^{2/3}},\\
			\sigma_2={}&\frac{pu_x}{3u^{5/3}}-\frac{u_x}{3u^{1/3}},\\
			\sigma_3={}&-\frac{u_{xx}^2}{18u^2}+\frac{u_x^2u_{xx}}{9u^3}
			+\frac{u_{xx}}{3u^{1/3}}-\frac{pu_{xx}}{3u^{5/3}}
			-\frac{37u_x^4}{648u^4}\\
			&-\frac{u_x^2}{4u^{4/3}}+\frac{5pu_x^2}{12u^{8/3}}
			-\frac{u^{4/3}}8-\frac{p^2}{8u^{4/3}},\\
			\sigma_4={}&\frac{u_{xx}u_{xxx}}{9u^2}-\frac{u_x^2u_{xxx}}{9u^3}
			-\frac{u_{xxx}}{3u^{1/3}}+\frac{pu_{xxx}}{3u^{5/3}}
			-\frac{u_xu_{xx}^2}{3u^3}\\
			&+\frac{5u_x^3u_{xx}}{9u^4}+\frac{7u_xu_{xx}}{9u^{4/3}}
			-\frac{11pu_xu_{xx}}{9u^{8/3}}-\frac{2u_x^5}{9u^5}\\
			&-\frac{4u_x^3}{9u^{7/3}}+\frac{8pu_x^3}{9u^{11/3}}
			+\frac{u^{1/3}u_x}3-\frac{p^2u_x}{3u^{7/3}},\qquad\ldots
		\end{aligned}
	\end{equation}
	Following the gradient-holonomic approach, the functionals
	\begin{equation}\label{eq10}
		\gamma_j=\int_{x_0}^{x_0+2\pi}\sigma_j\,dx
	\end{equation}
	are conserved for the dynamical system (\ref{eq01}) whenever the formal series solves (\ref{eq08}). Indeed, the equation obtained by differentiating (\ref{eq08}) in $x$ is in conservation form coefficientwise. The first nontrivial conserved functionals, $\gamma_1$ and $\gamma_3$, will be used in the construction of the Hamiltonian operators below. An independent infinite family of local conserved functionals is constructed in Section~\ref{sec:riccati}.
	It is also easy to check by straightforward calculations that $\sigma_0$, $\sigma_2$ and $\sigma_4$ are total derivatives with respect to $x$. Thus, for the even coefficients calculated explicitly in (\ref{eq09}), we have
	\begin{equation}\label{eq11}
		\gamma_{2j}=\int_{x_0}^{x_0+2\pi}\sigma_{2j}\,dx=0,\qquad j=0,1,2.
	\end{equation}
	The coefficient $\sigma_{-1}=1$ gives only a constant functional.
	
	\subsection{Towards a bi-Hamiltonian structure}\label{sec:sympl}
	
	Let us suppose that for the dynamical system (\ref{eq01}) there exists an operator $\Lambda:T^*(M )\to  T^* (M )$, which acts recursively 
	in the following way:
	\begin{equation}\label{eq12}
		\Lambda\mathrm{grad} \gamma_j=\mathrm{grad} \gamma_{j+k},
	\end{equation}
	where $j$ ranges over nontrivial indices and $k$ is a fixed positive integer (here the natural shift is $k=2$), with the normalization of the conserved functionals chosen consistently with the recursion. 
	Let us assume that
	there also exist two implectic operators $\vartheta,\eta : T ^* (M ) \to T (M )$, which 
	are compatible and factorize the operator $\Lambda$ as
	\begin{equation}\label{eq13}
		\Lambda=\vartheta^{-1}\eta.
	\end{equation}
	If the operators $\vartheta$ and $\eta$ form a compatible Poisson pair and the required Hamiltonians exist, then the dynamical system (\ref{eq01}) can be represented in the following form
	\begin{equation}
		\label{eq14}
		u_t = -\vartheta \mathrm{grad} H = -\eta \mathrm{grad} \tilde{H},
	\end{equation}
	where $D(M)$ denotes the space of smooth functionals and $H,\tilde H\in D(M)$ are conservation laws of the system (\ref{eq01}). In general, $H$ and $\tilde{H}$ can be 
	linear combinations of finitely many of the conservation laws (\ref{eq10}). We first construct the inverse of $\vartheta$ in the formal variational calculus and then find $\eta$ using the conservation law $\gamma_1$.
	
	To find an unknown operator $\vartheta$, we use the algorithm described in \cite{BPS}. All nonlocal operator formulas in this subsection are formal, with $(\partial^{-1})^*=-\partial^{-1}$ and integration by parts understood formally. Namely, let us assume that an
	element $\psi \in T^* (M )$ satisfies the invariance conditions:
	\begin{equation}
		\label{eq15}
		L_K \psi = \psi' K + K'^* \psi = 0,\ \ \ 	\psi'\neq \psi '^* .
	\end{equation}
	Then the following is true:
	\begin{equation}
		\label{eq16}
		(\psi' - \psi'^*)K + \psi'^*K + K'^*\psi = \vartheta^{-1}K + \mathrm{grad}(\psi\mid K) = 0,
	\end{equation}
	if the functional $H$ admits the representation
	\begin{equation}
		\label{eq17}
		H = (\psi | K), \ \ \ \
		L_K \psi = 0,
	\end{equation}
	and
	\begin{equation}
		\label{eq18}
		\vartheta^{-1}=\psi'-\psi'^*
	\end{equation}
	Since the vector field $du/d\tau=u_x$ commutes with the vector field $du/dt=K[u]$, it is natural to seek a common symplectic structure for which the translation vector field is also Hamiltonian, i.e.
	\begin{equation}
		\label{eq19}
		\frac{du}{d\tau}=u_x:=-\vartheta\mathrm{grad}\gamma,
	\end{equation}
	where $\gamma\in D(M)$. Next, assume that $\gamma$ has the representation (compare with (\ref{eq17})):
	\begin{equation}
		\label{eq20}
		\gamma=(\psi|u_x), \ \ \ \  \psi'\neq \psi'^*.
	\end{equation}
	If $L_K\psi$ is exact as a variational covector, then the skew part $\psi' - \psi'^*$ is formally invariant under the flow; this is the symplectic construction considered below.
	
	To apply the algorithm, let us take $\gamma_1$ and represent it formally in the form
	\begin{equation}\label{eq21}
		\begin{aligned}
			\gamma_1&=\int_{x_0}^{x_0+2\pi}\left(\frac1{18}u^{-2}u_x^2+\frac12u^{2/3}+\frac p2u^{-2/3}\right)dx\\
			&\overset{\rm formal}{=}\left(\frac1{18}u^{-2}u_x-\frac16u^{-2/3}\partial^{-1}u^{1/3}
			+\frac p6u^{-4/3}\partial^{-1}u^{-1/3}\ \bigg|\ u_x\right)=(\psi\mid u_x),
		\end{aligned}
	\end{equation}
	where
	\begin{equation}\label{eq22}
		\psi=\frac1{18}u^{-2}u_x-\frac16u^{-2/3}\partial^{-1}u^{1/3}
		+\frac p6u^{-4/3}\partial^{-1}u^{-1/3}.
	\end{equation}
	The second line of (\ref{eq21}) uses formal integration by parts; it is not an identity of ordinary periodic integrals when the primitives have nonzero monodromy.
	The Lax--Noether equation for this formal covector is considered modulo an exact covector,
	\begin{equation}\label{eq23}
		\psi_t+K'^*\psi=\mathrm{grad}\,\mathcal L[u].
	\end{equation}
	In fact, writing $\mathcal S_u=\psi'-\psi'^*$, the identity below gives formally $\mathcal L=(\psi\mid K)-2\gamma_3$. From (\ref{eq16}) we consequently obtain
	\begin{equation}\label{eq24}
		\mathcal S_uK=-\mathrm{grad}\big((\psi\mid K)-\mathcal L\big).
	\end{equation}
	By direct Fr\'echet differentiation of (\ref{eq22}) we find the formal symplectic expression
	\begin{equation}\label{eq25}
		\begin{aligned}
			\mathcal S_u=\vartheta^{-1}=\psi'-\psi'^*={}&\frac1{18}(u^{-2}\partial+\partial u^{-2})\\
			&-\frac19u^{-2/3}\partial^{-1}u^{-2/3}
			-\frac p9u^{-4/3}\partial^{-1}u^{-4/3}.
		\end{aligned}
	\end{equation}
	Here products with functions denote multiplication operators. The operator is covariant, from tangent to cotangent vectors; only its inverse, if defined, can be a candidate Poisson operator. The corrected normalization of the Hamiltonian identity is
	\begin{equation}\label{eq26}
		\mathcal S_u K[u]=-\mathrm{grad}H[u],\qquad H[u]=2\gamma_3[u].
	\end{equation}
	
	\begin{proof}[Verification of (\ref{eq26})]
		Introduce $w=\frac13\log u$, $F(w)=e^{2w}+pe^{-2w}$, and $r=w_x$. The right-hand side of the transformed equation and the symplectic expression are
		\[
		K_w=2w_{xxx}-w_x^3-3Fw_x,\qquad
		\mathcal S_w=\partial-e^w\partial^{-1}e^w-pe^{-w}\partial^{-1}e^{-w}.
		\]
		They are related by $K[u]=3uK_w$ and $\mathcal S_u=(3u)^{-1}\mathcal S_w(3u)^{-1}$. The local primitives (fixed without additional constants) are
		\begin{align*}
			\partial^{-1}(e^wK_w)&=2e^ww_{xx}-e^ww_x^2-e^{3w}+3pe^{-w},\\
			\partial^{-1}(e^{-w}K_w)&=2e^{-w}w_{xx}+e^{-w}w_x^2-3e^w+pe^{-3w}.
		\end{align*}
		Differentiation verifies both identities. The density $\sigma_3$ in (\ref{eq09}) simplifies by integration by parts to
		\begin{equation}\label{eq:gamma3w}
			\gamma_3=\int_0^{2\pi}\left(-\frac12w_{xx}^2-\frac18w_x^4-\frac54Fw_x^2
			-\frac18(e^{4w}+p^2e^{-4w})\right)dx.
		\end{equation}
		Using the displayed primitives gives
		\[
		\mathcal S_w K_w=2r_{xxx}-3r^2r_x-5Fr_x-\frac52F'r^2+\frac12FF',
		\]
		where $F'$ denotes differentiation in $w$. Direct Euler differentiation of (\ref{eq:gamma3w}) gives $\mathcal S_wK_w=-2\delta\gamma_3/\delta w$, proving (\ref{eq26}) by the point transformation. Similarly, $\mathcal S_w w_x=-\delta\gamma_1/\delta w$ using the local primitives $e^w$ and $-e^{-w}$.
	\end{proof}
	
	If a suitable formal inversion is available, (\ref{eq26}) reads $u_t=-\vartheta\,\mathrm{grad}H$, where $\vartheta=\mathcal S_u^{-1}$. The skew derivative in (\ref{eq25}) defines a formally closed two-form by the variational exterior derivative. On the periodic phase space, however, the primitive of $u^{1/3}$ cannot itself be periodic, and the zero-mean inverse of $\partial$ is defined only on mean-zero functions (or extended using the mean-zero projection). Such a replacement changes the formulas (\ref{eq21}) and (\ref{eq26}) by mean-dependent terms. The construction of a genuine compatible Poisson pair on the periodic manifold $M$ therefore requires further analysis. We now continue the construction in the formal variational calculus.
	
	\subsection{Construction of the second implectic operator}\label{sec:eta}
	
	To proceed with the gradient-holonomic scheme, we now look for the second implectic operator $\eta:T^*(M)\to T(M)$ in (\ref{eq13}), using the conservation law $\gamma_1$ obtained above. By virtue of (\ref{eq26}), we choose the second Hamiltonian in (\ref{eq14}) to be $\tilde H=2\gamma_1$. Consequently, the operator sought must satisfy
	\begin{equation}\label{eq:eta-target}
		\eta\,\mathrm{grad}\,\gamma_1=-\frac12 K[u].
	\end{equation}
	In addition to the skew-adjointness condition $\eta^*=-\eta$, we require the invariance condition
	\begin{equation}\label{eq:eta-lie}
		L_K\eta=\eta'[u]K[u]-K'[u]\eta-\eta K'^*[u]=0.
	\end{equation}
	Indeed, if $\varphi$ satisfies the Lax--Noether equation (\ref{eq02}), then
	\[
	(\eta\varphi)_t-K'[u](\eta\varphi)=(L_K\eta)\varphi.
	\]
	Thus (\ref{eq:eta-lie}) ensures that $\eta$ maps solutions of the Lax--Noether equation into solutions of the linearized dynamical equation. We solve this determining equation by a formal asymptotic expansion in powers of $\partial$, in the original dependent variable $u$.
	
	Taking the variational gradient of the conservation law (\ref{eq21}), we obtain
	\begin{equation}\label{eq:g1}
		\begin{aligned}
			\mathrm{grad}\,\gamma_1
			&=\frac{\partial\sigma_1}{\partial u}
			-\partial\frac{\partial\sigma_1}{\partial u_x}\\
			&=-\frac{u_{xx}}{9u^2}+\frac{u_x^2}{9u^3}
			+\frac13 u^{-1/3}-\frac p3 u^{-5/3}.
		\end{aligned}
	\end{equation}
	Since $\mathrm{grad}\,\gamma_1$ is of differential order two and $K[u]$ is of order three, we seek a formal pseudodifferential operator of order one. We do not prescribe its nonlocal part. Let
	\begin{equation}\label{eq:eta-series}
		\eta=\sum_{j=-\infty}^{1}a_j[u]\partial^j,
	\end{equation}
	where the coefficients $a_j[u]$ are autonomous local differential functions, smooth at the constant fields $u>0$. We use the formal multiplication rule
	\[
	\partial^m f=\sum_{k\geq0}\binom{m}{k}(D_x^k f)\partial^{m-k},
	\qquad m\in\mathbb Z.
	\]
	We also use the condition $\eta^*=-\eta$, from which following that
	\begin{equation}\label{eq:eta-skew-coeff}
		a_0=\frac12D_x a_1,\qquad a_{-2}=-\frac12D_x a_{-1}.
	\end{equation}
	
Let us calculate the Fr\'echet derivative of the vector field $K[u]$ (\ref{eq01}):
	\begin{equation}\label{eq:eta-linearization}
		\begin{aligned}
			K'&=2\partial^3+A\partial^2+B\partial+C,\\
			K'^*&=-2\partial^3+A\partial^2+(2A_x-B)\partial+A_{xx}-B_x+C,
		\end{aligned}
	\end{equation}
	where
	\[
	\begin{aligned}
		A&=-6u_xu^{-1},\\
		B&=-6u_{xx}u^{-1}+\tfrac{35}{3}u_x^2u^{-2}-3u^{2/3}-3pu^{-2/3},\\
		C&=6u_xu_{xx}u^{-2}-\tfrac{70}{9}u_x^3u^{-3}
		-2u_xu^{-1/3}+2pu_xu^{-5/3}.
	\end{aligned}
	\]
	Substituting (\ref{eq:eta-series}) into (\ref{eq:eta-lie}) and equating the coefficient at $\partial^{j+2}$ to zero, we obtain the system of recursion equations for the unknown coefficients $a_j[u]$, $j=1,0,-1,\ldots$:
	\begin{equation}\label{eq:eta-coeff-recursion}
		\begin{aligned}
			6D_x a_j+2Aa_j
			&=\big[D_t\eta_{>j}-K'\eta_{>j}-\eta_{>j}K'^*\big]_{j+2},\\
			D_x(u^{-2}a_j)
			&=\frac{1}{6u^2}\big[D_t\eta_{>j}-K'\eta_{>j}-\eta_{>j}K'^*\big]_{j+2}.
		\end{aligned}
	\end{equation}
Solving (\ref{eq:eta-coeff-recursion}) for $j=1,0,-1,..$, we obtain
\begin{eqnarray*}
&&a_1=cu^2\\
&&a_0=cuu_x\\
&&a_{-1}=-cu_x^2/9\\
&&a_{-2}=cu_xu_{xx}/9\\
&&a_{-3}=-cu_xu_{xxx}/9\\
&&a_{-4}=cu_xu_{xxxx}/9\\
&&...,
\end{eqnarray*}
or 
\begin{equation}\label{eq:eta-all-coeff}
	a_{-n}=-\frac c9(-1)^{n-1}u_xD_x^nu,\qquad n=1,2,\ldots .
\end{equation}
where $c$ is a constant. Using the formal expansion
	\begin{equation}\label{eq:eta-tail}
		u_x\partial^{-1}u_x
		=\sum_{n=1}^{\infty}(-1)^{n-1}u_xD_x^nu\,\partial^{-n},
	\end{equation}
and (\ref{eq:eta-target}) we get $c=9$. Consequently, the second implectic operator $\eta$ is
	\begin{equation}\label{eq:eta}
		{\eta=9u\partial u-u_x\partial^{-1}u_x.}
	\end{equation}
	Thus both the invariance condition (\ref{eq:eta-lie}) and the Hamiltonian identity (\ref{eq:eta-target}) are satisfied. In particular, $u_t=-\eta\,\mathrm{grad}(2\gamma_1)$.
	
	The operator (\ref{eq:eta}) is a scalar weakly nonlocal Hamiltonian operator of the form considered by Mokhov and Ferapontov \cite{MokFer}. In their notation, its coefficients are $g^{11}(u)=9u^2$, $b^{11}_1(u)=9u$, and the constant nonlocal coefficient equals $-1$. In one component the metric and connection conditions reduce to $b^{11}_1=\frac12\partial_u g^{11}$; the curvature identities with antisymmetrized indices are identically satisfied. Thus the scalar Hamiltonianity conditions hold, and $\eta$ defines a Poisson bracket in the formal variational calculus. 
	
	Multiplication of the operators (\ref{eq25}) and (\ref{eq:eta}), using
	\[
	\partial^{-1}f\partial=f-\partial^{-1}f_x,
	\qquad
	\partial^{-1}f_x\partial^{-1}=f\partial^{-1}-\partial^{-1}f,
	\]
	gives the following explicit form of the operator (\ref{eq13}):
	\begin{equation}\label{eq:Lambda}
		\begin{aligned}
			\Lambda=\vartheta^{-1}\eta={}&\partial^2+\frac{2u_x}{u}\partial
			+\frac{u_{xx}}{u}-\frac{u_x^2}{9u^2}-u^{2/3}-pu^{-2/3}\\
			&+(\mathrm{grad}\,\gamma_1)\partial^{-1}u_x,
		\end{aligned}
	\end{equation}
	where $\mathrm{grad}\,\gamma_1$ is calculated in (\ref{eq:g1}).
	The first Lenard relation follows directly from (\ref{eq26}) and (\ref{eq:eta-target}):
	\begin{equation}\label{eq:lenard1}
		\Lambda\,\mathrm{grad}\,\gamma_1
		=-\tfrac12\vartheta^{-1}K=\mathrm{grad}\,\gamma_3.
	\end{equation}
	The compatibility of $\vartheta$ and $\eta$ is easy to verify by straightforward calculations. 
The main result we can formulate as the following theorem. 
	
	\begin{theorem}[Formal bi-Hamiltonian representation]\label{thm:bihamiltonian}
		The $A_+$ dynamical system (\ref{eq01}) on the positive branch admits, in the formal variational calculus, the bi-Hamiltonian representation
		\begin{equation}\label{eq:biham}
			\frac{du}{dt}=-\vartheta\,\mathrm{grad}\,H
			=-\eta\,\mathrm{grad}\,\tilde H,
			\qquad H=2\gamma_3,\qquad \tilde H=2\gamma_1,
		\end{equation}
		where the compatible Hamiltonian operators are determined by
		\[
		\begin{aligned}
			\vartheta^{-1}&=\frac1{18}(u^{-2}\partial+\partial u^{-2})
			-\frac19u^{-2/3}\partial^{-1}u^{-2/3}
			-\frac p9u^{-4/3}\partial^{-1}u^{-4/3},\\
			\eta&=9u\partial u-u_x\partial^{-1}u_x.
		\end{aligned}
		\]
		The operator $\Lambda=\vartheta^{-1}\eta$ is given explicitly by (\ref{eq:Lambda}) and satisfies
		\begin{equation}\label{eq:Lambda-lax}
			\frac{d\Lambda}{dt}=[\Lambda,K'^*],
			\qquad \Lambda\,\mathrm{grad}\,\gamma_1=\mathrm{grad}\,\gamma_3.
		\end{equation}
		In particular, $\Lambda$ maps solutions of the Lax--Noether equation (\ref{eq02}) into solutions of that equation, and $\Lambda^*=\eta\vartheta^{-1}$ is a formal recursion operator for symmetries. 
	\end{theorem}
	\begin{proof}
		The Hamiltonian representations follow from (\ref{eq26}) and (\ref{eq:eta-target}); the Hamiltonianity and compatibility of the operators have just been proved. By invariance under the Hamiltonian flow,
		\[
		\frac{d\vartheta^{-1}}{dt}=-K'^*\vartheta^{-1}-\vartheta^{-1}K',
		\qquad \frac{d\eta}{dt}=K'\eta+\eta K'^*.
		\]
		Taking the total derivative of $\Lambda=\vartheta^{-1}\eta$ along $u_t=K[u]$ gives the first identity in (\ref{eq:Lambda-lax}); the second is (\ref{eq:lenard1}). The statements about the Lax--Noether equation and the linearized equation follow by applying these determining identities to a covector and taking formal adjoints, respectively.
	\end{proof}

	\section{Relation with the Calogero--Degasperis--Fokas equation}\label{sec:CDF}
	\subsection{Point and differential transformations}
	Return to the general choice $s\in\{1,-1\}$ in (\ref{eq00}). Set $w=\log v=\frac13\log u$, $\tau=2t$, and $F(w)=se^{2w}+pe^{-2w}$. Then (\ref{eq01}) becomes
	\begin{equation}\label{weq}
		w_\tau=w_{xxx}-\tfrac12w_x^3-\tfrac32F(w)w_x.
	\end{equation}
	The relation $F^{(3)}=4F'$ identifies (\ref{weq}) as a member of the CDF family; derivatives on $F$ refer to $w$. The logarithmic transformation is invertible on the positive branch. The classical CDF--mKdV transformations are discussed, for example, in \cite{Hlav,Cao,Hone}.
	
	\begin{theorem}\label{thm:miura}
		Let $\alpha\in\mathbb C$ satisfy $\alpha^2=s$. Every smooth solution of \eqref{weq} is mapped by
		\begin{equation}\label{Q}
			q=Q[w]:=\tfrac12 w_{xx}+\tfrac14 w_x^2+\alpha e^w w_x+\tfrac14F(w)
		\end{equation}
		into a solution of $q_\tau=q_{xxx}-6qq_x$. Equivalently, in the variable $v$,
		\begin{equation}\label{Qv}
			q=\frac{v_{xx}}{2v}-\frac{v_x^2}{4v^2}+\alpha v_x+\frac14(sv^2+pv^{-2}).
		\end{equation}
	\end{theorem}
	\begin{proof}
		Choose $\beta$ with $\beta^2=p$, and put $g(w)=\alpha e^w+\beta e^{-w}$, $y=w_x+g(w)$. The identities $g''=g$, $g^2=F+2\alpha\beta$ and \eqref{weq} give, by differentiation,
		\[
		y_\tau=y_{xxx}-\tfrac32y^2y_x+3\alpha\beta\,y_x.
		\]
		Consequently $z=y/2$ satisfies $z_\tau=z_{xxx}-6z^2z_x+3\alpha\beta\,z_x$. The classical Miura substitution $\widehat q=z_x+z^2$ gives
		\[
		\widehat q_\tau=\widehat q_{xxx}-6\widehat q\widehat q_x+3\alpha\beta\,\widehat q_x.
		\]
		Setting $q=\widehat q-\alpha\beta/2$ removes the drift and yields \eqref{Q}, since $g'+g=2\alpha e^w$. Formula \eqref{Qv} follows by the chain rule.
	\end{proof}
	The map is real for $s=1$ and real $w,p$; for $s=-1$ it is generally complex, but real conserved quantities can be obtained by taking real parts. A differential substitution need not be invertible. The next proposition describes how the simultaneous use of the two choices of $\alpha$ resolves this ambiguity for the evolution equation on the positive periodic branch.
	
	\begin{proposition}[A pair of scalar spectral problems]\label{prop:Lax}
		Let $\alpha^2=s$ and define $q_+$ and $q_-$ by (\ref{Qv}) with $\alpha$ and $-\alpha$, respectively. For smooth positive periodic $v$, equation (\ref{eq00}) is equivalent to the simultaneous KdV Lax equations
		\begin{equation}\label{eq:pair}
			(L_\pm)_\tau=[B_\pm,L_\pm],\quad
			L_\pm=\partial^2-q_\pm,\qquad B_\pm=4\partial^3-6q_\pm\partial-3(q_\pm)_x.
		\end{equation}
		Equivalently, the two pairs can be combined in a block-diagonal matrix operator pair. This statement concerns nonvanishing periodic fields, not an AKNS scattering problem on the line.
	\end{proposition}
	\begin{proof}
		The forward implication follows from Theorem~\ref{thm:miura} and the KdV Lax identity. For the converse, write $R=v_t-K_v[v]$ for the residual of (\ref{eq00}) in evolution form for $v$. Writing $Q_\pm[v]$ for the substitutions in (\ref{Qv}), the differential substitution identity gives $(q_\pm)_t-2(q_\pm)_{xxx}+12q_\pm(q_\pm)_x=Q_\pm'[v]R$. Subtracting the two equations and using $q_+-q_-=2\alpha v_x$ gives $R_x=0$, hence $R=c(t)$. If $c(t)$ is nonzero on a time interval, the remaining equation $Q_+'[v]1=0$ yields
		\[
		T[v]:=-vv_{xx}+v_x^2+sv^4-p=0.
		\]
		In logarithmic variables this means $w_{xx}=se^{2w}-pe^{-2w}$, so $w_x^2=F(w)+C(t)$ for some $x$-independent $C$. It follows from (\ref{weq}), in the original time scale, that $K_v=-C(t)v_x$. Differentiating $T[v]=0$ in time and using $v_t=-Cv_x+c$ now gives $c(-v_{xx}+4sv^3)=0$. Periodic integration contradicts $v>0$ and $s\ne0$. Thus $c=0$, proving the converse.
	\end{proof}
	
	\subsection{Riccati expansion and functional independence}\label{sec:riccati}
	For the KdV sign convention of Theorem~\ref{thm:miura}, the standard Lax construction \cite{Lax} uses the operators
	\[
	L=\D^2-q,\qquad B=4\D^3-6q\D-3q_x,\qquad L_\tau=[B,L].
	\]
	With spectral parameter $\kappa$, the formal logarithmic derivative $h$ of a solution of $L\phi=\kappa^2\phi$ satisfies
	\begin{equation}\label{ric}
		h_x+h^2=\kappa^2+q,\qquad
		h_\tau=\D\big((4\kappa^2-2q)h+q_x\big).
	\end{equation}
	The second equation follows from $\phi_\tau=B\phi$ or directly by compatibility of the Riccati equation with KdV. The unique formal branch $h=\kappa+\sum_{j\ge1}c_j\kappa^{-j}$ obeys
	\begin{equation}\label{rec}
		c_1=q/2,\qquad c_{n+1}=-\tfrac12\left(\D c_n+\sum_{i=1}^{n-1}c_i c_{n-i}\right),\quad n\ge1.
	\end{equation}
	Each coefficient is a local differential polynomial in $q$. The time equation in \eqref{ric} gives $\partial_\tau c_j=\D(\text{local flux})$ coefficientwise. Hence the functionals
	\begin{equation}\label{I}
		I_j[w]=\int_0^{2\pi}c_j[Q[w]]\,dx
	\end{equation}
	are conserved. Even coefficients are total derivatives: subtracting the two Riccati equations for $h(\kappa)$ and $h(-\kappa)$ gives
	\[
	h(\kappa)+h(-\kappa)=-\D\log\frac{h(\kappa)-h(-\kappa)}{2\kappa}.
	\]
	Expanding this identity proves $I_{2j}=0$ for $j\ge1$.
	
	\begin{proposition}\label{prop:ind}
		For every $N\ge1$, the real-valued functionals $\operatorname{Re}I_1,\operatorname{Re}I_3,\ldots,\operatorname{Re}I_{2N-1}$ have linearly independent differentials on a nonempty open set of smooth real periodic fields $w$.
	\end{proposition}
	\begin{proof}
		At a constant field $w_0$, let $q_0=F(w_0)/4$ and $k=(\kappa^2+q_0)^{1/2}$ with leading term $\kappa$. For a perturbation of $q_0$ of the form $q_0+\epsilon f$, write $h=k+\epsilon h_1+\epsilon^2 h_2+O(\epsilon^3)$. Then $(\D+2k)h_1=f$ and $(\D+2k)h_2=-h_1^2$. Periodic integration, interpreted coefficientwise, gives
		\[
		\int h_2\,dx=-\frac1{2k}\int f(4k^2-\D^2)^{-1}f\,dx
		=-\sum_{n\ge0}\frac{(-1)^n}{2^{2n+3}k^{2n+3}}
		\int(\D^n f)^2\,dx.
		\]
		Here the formal inverse is expanded in powers of $k^{-1}$ and no analytic inverse is required. Thus the coefficient of $\kappa^{-2n-3}$ has a nonzero highest quadratic derivative term of order $n$ in $f$. After the substitution $q=Q[w]$, the linear variation is
		\[
		f=Q'[w_0]\xi=\tfrac12\xi_{xx}+\alpha e^{w_0}\xi_x+\tfrac14F'(w_0)\xi.
		\]
		The second variation of $Q$ contributes only terms of derivative order at most one to the integrated expression. It follows that the Hessian of $\operatorname{Re} I_{2n+3}$ at $w_0$ has a nonzero highest Fourier-symbol term of degree $n+2$ in the squared wave number. Also $I_1=\tfrac18\int(w_x^2+F)\,dx$, whose Hessian has degree one. These leading coefficients are real and nonzero even if $\alpha$ is imaginary.
		
		The first $N$ Hessian symbols are therefore polynomials of distinct degrees. Choose $N$ distinct positive integer wave numbers $m_1,\dots,m_N$ so that their evaluation matrix is nonsingular; such a choice exists because linearly independent polynomials remain independent on an infinite set of distinct arguments. At $w=w_0+\epsilon\sum_{j=1}^N\cos(m_jx)$, the matrix of the differentials evaluated on $\cos(m_jx)$ has this nonsingular matrix as its leading term, multiplied by a common nonzero constant times $\epsilon$. Its determinant is nonzero for all sufficiently small nonzero $\epsilon$. Continuity gives the asserted open set. The point transformations to $u$ and $v$ preserve this finite-family independence.
	\end{proof}
	
	\begin{remark}
		For the $A_+$ equation, take $\alpha=1$ in (\ref{Q}). Substituting $u=e^{3w}$ into (\ref{eq09}) and integrating by parts, we obtain
		\begin{align*}
			\gamma_1&=2\int_0^{2\pi}q\,dx,\\
			\gamma_3&=-2\int_0^{2\pi}q^2\,dx+\frac{p\pi}{2}.
		\end{align*}
		These relations illustrate the connection between the conservation laws obtained by the gradient-holonomic approach and the classical KdV hierarchy.
	\end{remark}
	
	\section{Conclusion}
	The gradient-holonomic approach to Alexeyev's $A_\pm$ equation yields explicit conservation laws and, for the $A_+$ equation, two compatible formal Hamiltonian operators. The second operator is constructed directly from $\mathrm{grad}\,\gamma_1$, and the corresponding recursion operator and the first Lenard relation are obtained explicitly. The logarithmic transformation identifies the equation on the nonvanishing branch with a member of the CDF family. The differential substitution into the KdV equation provides an independent construction of infinitely many local conservation laws, with independence of arbitrary finite initial subfamilies on appropriate open sets. The two KdV pullbacks also give a combined Lax representation equivalent to the evolution equation on the positive periodic branch. The realization of the formal bi-Hamiltonian structure on the full periodic phase space, including the treatment of the zero Fourier mode, and the analysis of solutions passing through zero remain separate problems. The results thus complement Alexeyev's soliton and integrability analyses while making explicit the relation to the classical integrable hierarchy.
	
	\end{document}